\documentclass[graybox, natbib]{svmult}
\bibpunct{(}{)}{;}{a}{}{,} 

\usepackage{array}

\usepackage{doi}
\usepackage{mathptmx}       
\usepackage{helvet}         
\usepackage{courier}        
\usepackage{type1cm}        

\usepackage{makeidx}         
\usepackage{graphicx}        
\usepackage{multicol}        
\usepackage[bottom]{footmisc}
\usepackage[normalem]{ulem}	
\usepackage{hyperref}  
\usepackage{soul}   

\makeindex             

\usepackage[utf8]{inputenc}
\usepackage{subfigure}
\usepackage{amsmath,amssymb}
\usepackage{xspace}
\usepackage{miller}
\newcommand{\ie}{\textit{i.e.}\@\xspace}
\newcommand{\cf}{\textit{cf}.\@\xspace}

\newcommand{\abinitio}{\textit{ab initio}\@\xspace}
\newcommand{\Abinitio}{\textit{Ab initio}\@\xspace}
\newcommand{\apriori}{\textit{a priori}\@\xspace}

\newcommand{\ud}{\mathrm{d}}
\DeclareMathOperator{\Max}{Max}

\begin{document}

\title*{Ab Initio Models of Dislocations\thanks{To appear in
\emph{Handbook of Materials Modeling},
W. Andreoni and S. Yips (Eds.)}
}
\author{Emmanuel Clouet}

\institute{Emmanuel Clouet 
\at DEN-Service de Recherches de Métallurgie Physique, 
CEA, Université Paris-Saclay, 
F-91191 Gif-sur-Yvette, France,
\email{emmanuel.clouet@cea.fr}}

\maketitle

\abstract{This chapter reviews the different methodological aspects 
of the \abinitio modeling of dislocations. 
Such simulations are now frequently used to study the dislocation core, 
\ie the region in the immediate vicinity of the line defect 
where the crystal is so strongly distorted that an atomic description is needed.
This core region controls 
some dislocation fundamental properties, like their ability to glide 
in different crystallographic planes.
\Abinitio calculations based on the density functional theory 
offer a predictive way to model this core region.
Because dislocations break the periodicity of the crystal 
and induce long range elastic fields, 
several specific approaches relying on different boundary conditions 
have been developed to allow for the atomistic modeling of these defects 
in simulation cells having a size compatible with \abinitio calculations. 
We describe these different approaches which 
can be used to study dislocations with \abinitio calculations
and introduce the different analyses which are currently performed 
to characterize the core structure, 
before discussing how meaningful energy properties can be extracted 
from such simulations.}

\section{Introduction}

Dislocations are line defects which control the development of the plastic deformation
in crystals.  
These defects induce a long range stress field, which is well described by elasticity, and
dislocation elasticity theory offers a powerful framework to model dislocations 
and their interaction with their surrounding environment \citep{Hirth1982,Bacon1980}.
But some of their fundamental properties, like their glide plane and their mobility, 
highly depend on their core, \ie the region in the immediate vicinity of the defect
where the perturbation of the crystal is too important to be described by elasticity. 
The modeling of this core region necessitates an atomic description and  
atomistic simulations have thus become a valuable tool to study dislocation properties. 
Among such simulations, \abinitio calculations based on the density functional theory (DFT),
as they rely on an electronic description 
of the atomic bonding, appear as the most accurate and predictive.  
But as these calculations are still limited in the size of the system they can handle, 
typically at most a few hundred atoms, the \abinitio modeling of dislocations 
need special attention.  
Specific methodologies have been therefore developed to study dislocation core properties 
with \abinitio calculations.  
The purpose of this chapter is to review the different modeling approaches
for the \abinitio study of dislocations, 
starting from a quick overview of DFT formalism, 
before describing more thoroughly boundary conditions specific to dislocation models, 
then the analysis of the atomic structure in the dislocation core 
and finally the extraction of meaningful energy properties.
Beyond the examples illustrated in this chapter,
results which have been obtained from such \abinitio studies
for the dislocation core properties in different metals and semi-conductors 
can be found in the recent review of \cite{Rodney2017}.

\section{Ab initio calculations}

\Abinitio calculations describe the bonding between atoms 
thanks to the resolutions of the Schrödinger equation 
for the electrons of the system.
These are first-principles approaches as they do not use any experimental data 
and allows the modeling of atomic interaction only from the atomic number and other fundamental quantities.
Compared to empirical interatomic potentials, 
such approaches are completely transferable, without any parameterization depending on the environment
under study, but at the expense of a much higher CPU time.  
Although \abinitio in nature and usually very accurate, 
these approaches nevertheless rely on different approximations, 
the validity of which needs generally to be assessed. 

The most fundamental approximation is the Born-Oppenheimer approximation.  
As atom nuclei have a much higher mass than electrons, 
one can assume that the electrons are always equilibrated with respect to the positions of the nuclei
which are considered as immobile.  
The resolution of the Schrödinger equation for the electrons 
therefore leads to the energy of the system as a function of the atomic positions. 
Knowing this function and also its first derivatives, \ie the atomic forces, 
standard algorithms of atomic simulations can then be used. 
For the \abinitio modeling of dislocations, this is usually restricted 
to molecular statics, including energy barrier calculations, 
because of the high CPU burden of the energy and force calculation.

Most \abinitio calculations of dislocations are relying on the density functional theory
(DFT).  This makes use of the \cite{Hohenberg1964} theorem 
showing that the ground-state energy is the minimum of a functional 
depending only on the electronic density.  This dramatically simplifies the problem 
as the electronic density depends only on the position,
whereas the many-electron wave function entering Schrödinger equation 
is a function depending on the $3N$ electron coordinates, with $N$ the number of electrons in the system.
The \cite{Kohn1965} approach allows then a practical implementation,
where the Schrödinger equation is solved for an equivalent system of non-interacting electrons. 
This necessitates the definition of an unknown contribution to the Hamiltonian, 
the exchange and correlation potential. 
Most of dislocation calculations are performed with the local density (LDA) 
or the generalized gradient (GGA) approximations, assuming that this contribution 
depends only locally on the electronic density or also its gradient.

For dislocation calculations, it is enough to consider that only the electrons of the outer shells
participate to the atomic bonding. 
Electrons of the inner shells are not sensitive to the atom environment 
and can be assumed to have the same ground state as for the isolated atom. 
Kohn-Sham equations are then solved only for valence electrons. 
One can further reduce the CPU overhead by replacing with a pseudopotential
the interaction potential of the valence electrons with the ionic core. 
This pseudopotential aims to reduce the strong oscillations of the electronic wave functions 
close to the dislocation core, 
because the description of these oscillations necessitates a  large basis set,
while still leading to the correct wave functions outside this core region. 
Different pseudoization schemes, norm-conserving or ultrasoft pseudopotentials as well as 
the projected augmented wave (PAW) method, are available. 

\Abinitio codes used for dislocations are relying on Born-von Karman periodic 
boundary conditions to model the solid, whatever the boundary conditions used 
to incorporate a dislocation in the simulation cell (Section \ref{sec:boundary}). 
Electronic wave functions are thus a superposition of Bloch waves
with wavevectors spanning the first Brillouin zone. 
Integration in the reciprocal space is performed on a regular grid sampling 
the first Brillouin zone, using smearing functions to broaden the electronic density of states. 
Different basis sets can be used to describe the Bloch waves, 
with plane waves being the most popular choice for dislocations. 

\Abinitio approaches devoted to the study of dislocations are thus not specific: 
they are making use of standard DFT implementations which are now current modeling tools in solid state physics. 
Feature specific to dislocation modeling, as described in the next section, 
is the necessity to use a supercell large enough to let the dislocation core
adopt its fully relaxed configuration, 
with boundary conditions compatible
with the long range distortion induced by the defect. 
A high accuracy is also generally needed for such calculations 
as the energy variations involved by the dislocation core are usually small. 
For instance, the Peierls energy barrier opposing the glide of $1/2\,\hkl<111>$
screw dislocations in BCC transition metals does not exceed 100\,meV$/b$, 
where $b$, the norm of the Burgers vector, corresponds to the height of the simulation cell
necessary to model such a dislocation.

\section{Boundary conditions}
\label{sec:boundary}

The \abinitio modeling of dislocations needs special care in the way the boundary conditions 
are handled.  
First, a dislocation creates a long-range elastic field which needs to be taken into account.
Second, it is not possible to include a single dislocation in a simulation box 
with full periodic boundary conditions which usually constitute the paradigm 
in the modeling of bulk materials: the dislocation opens a displacement discontinuity
and another defect is needed to close the discontinuity and allow for periodicity.
As a result, different boundary conditions compatible with \abinitio calculations 
have been developed to model dislocations.

All approaches enforce periodicity in the direction of the dislocation line. 
In pure metals, one usually uses the shortest periodicity vector to define the dimension of the simulation cell 
in this direction, thus modeling an infinite straight dislocation. 
But this size needs to be increased if one wants to introduce a solute atom 
on the dislocation line, so as to minimize the interaction of the solute 
atom with its periodic images and truly study the interaction of the dislocation 
with a single foreign atom.
A larger size is also needed to model a kinked dislocation. 
This is usually possible only in covalent crystals where the atomic bonds are highly directional,
leading to abrupt kinks experiencing a non negligible energy barrier when migrating along the dislocation line.
In metallic systems with less directional atomic bonding, kinks are usually spread 
over a larger distance and are highly mobile, making it hard to stabilize them 
in a simulation cell whose size is compatible with \abinitio calculations. 

\subsection{Cluster approach}
\label{sec:cluster_fixed}

\begin{figure}[b]
	\begin{center}
		 \hspace*{1ex}\hfill
		(a)\includegraphics[width=0.3\linewidth]{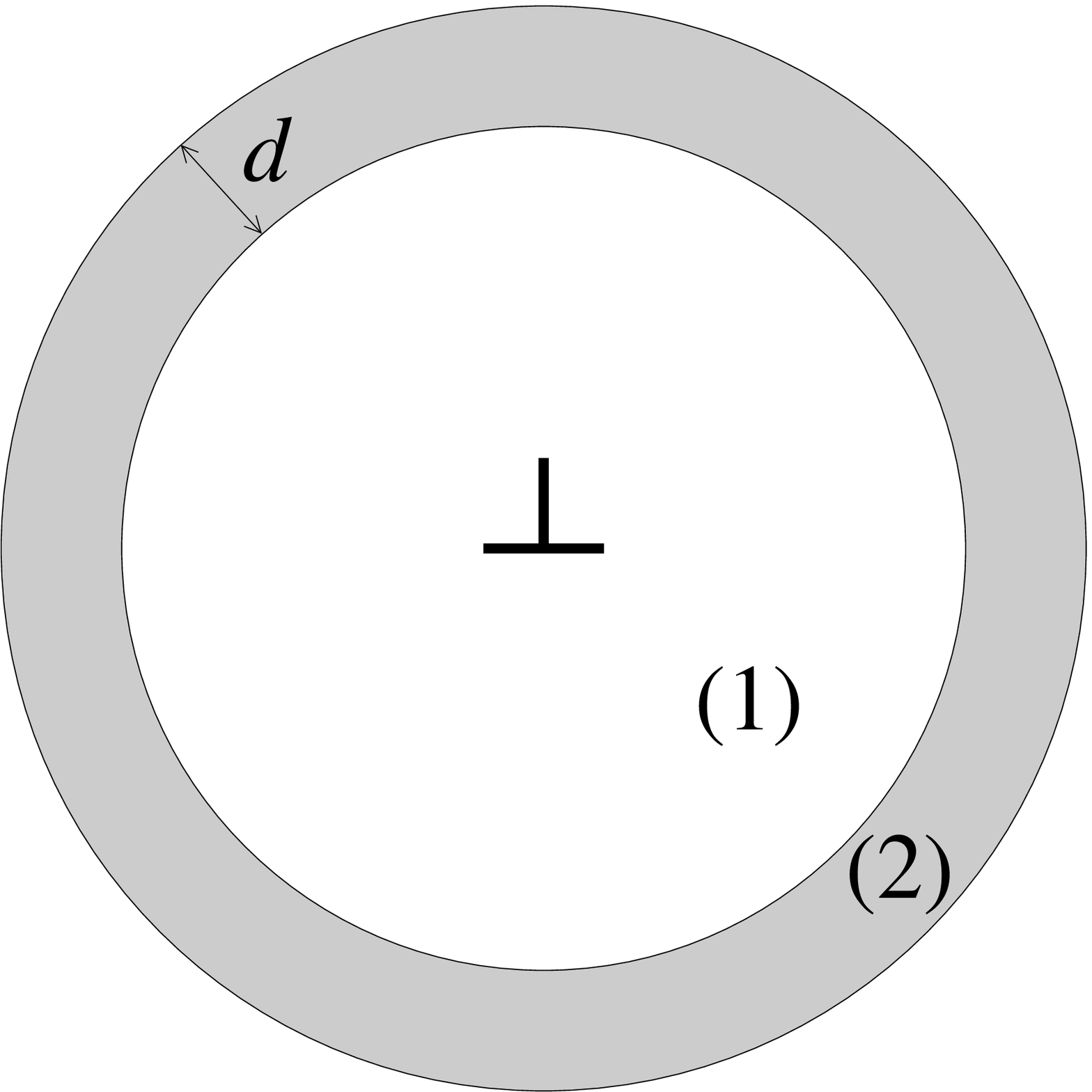}
		\hfill
		(b)\includegraphics[width=0.3\linewidth]{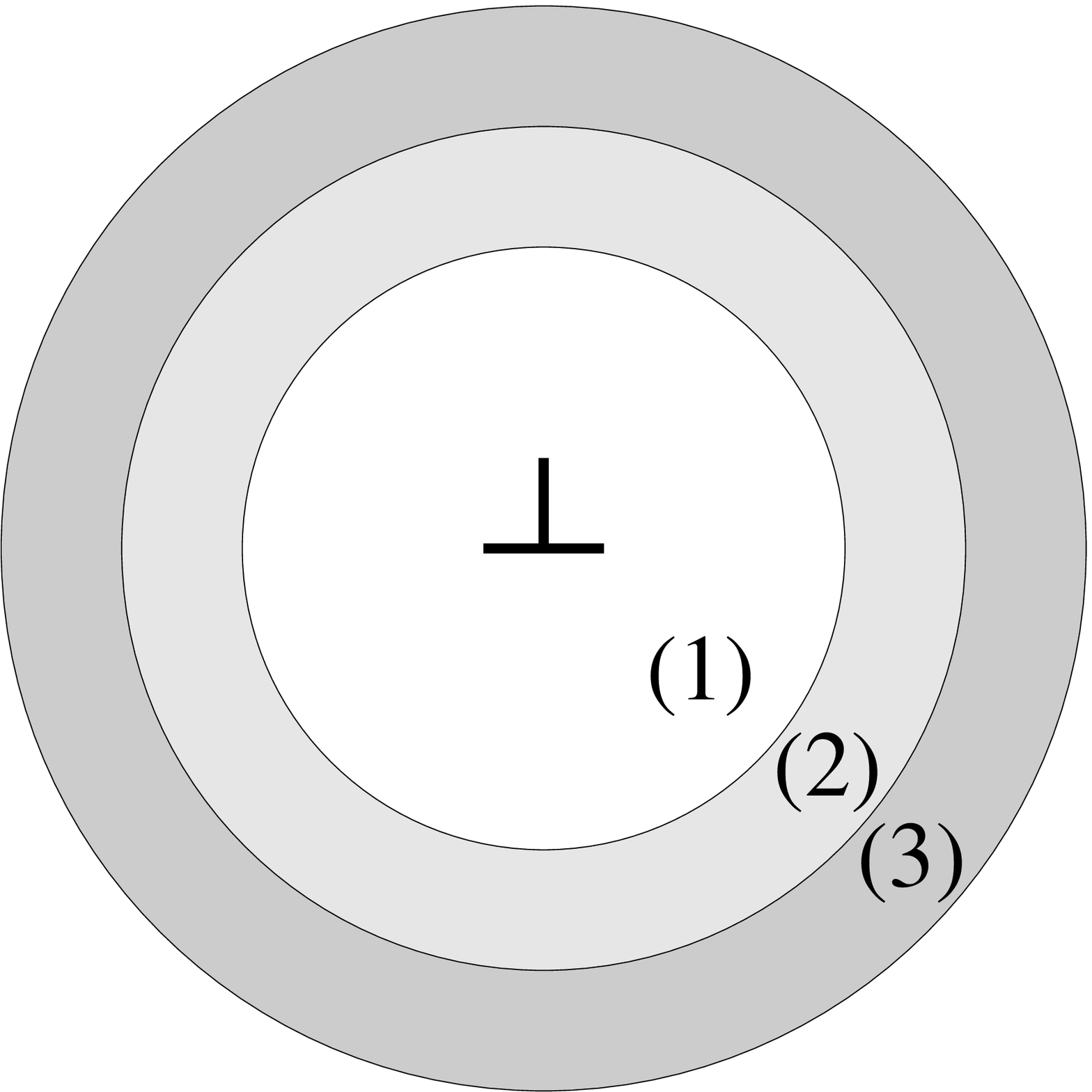}
		\hfill \hspace*{1ex}
	\end{center}
	\caption{Boundary conditions used to model an isolated straight dislocation 
	in the 	cluster approach.
	The outer boundary is either (a) rigid  
	or (b) flexible and controlled by lattice Green's functions
	or by coupling with an empirical potential.  }
	\label{fig:cluster_sketch}
\end{figure}

The easiest way to model a dislocation is to use an infinite cylinder
whose axis coincides with the dislocation line.
Periodicity is enforced only along the dislocation line.
The dislocation is created by displacing all atoms according 
to the Volterra solution given by anisotropic elasticity theory for the dislocation displacement field 
\citep{Stroh1958,Stroh1962}.
Atoms at the cylinder surface (region 2 in Fig. \ref{fig:cluster_sketch}a)
are kept fixed in their initial positions 
and only atoms inside the cylinder are relaxed.
One thus models an isolated dislocation in an infinite continuum.

But this modeling approach has severe drawbacks.
The elastic solution used to fix the atoms at the boundary is only approximate 
as it relies on linear elasticity, thus neglecting crystal anharmonicity 
which can be strong close to the dislocation line.
Moreover, the Volterra elastic solution, used to fix the atoms at the boundary, 
only corresponds to the long-range elastic field of the dislocation. 
Close to the dislocation line some additional contributions,
the dislocation core field, need to be accounted for \citep{Eshelby1953}. 
A spreading of the dislocation core can be the reason for the existence
of such a core field, but even dislocations with a compact core, like $\langle111\rangle$
screw dislocations in BCC metals, possess a non negligible core field. 
Although this core field decays more rapidly than the Volterra elastic field, 
the size of the simulation boxes that can be handled by \abinitio calculations 
are never large enough to neglect it.
The rigid boundary conditions do not allow the correct development of this core field 
and thus perturb the relaxation of the dislocation core.

The fixed atomic positions imposed at the boundary also prevent use of this method 
to determine the lattice friction opposing 
dislocation motion.
If the dislocation moves during the simulation, this boundary condition 
will not be compatible anymore with the new dislocation position. 
This induces indeed a back-stress opposing the dislocation motion. 
As a result, any simulation relying on this boundary condition will overestimate 
the dislocation Peierls stress, which is the minimum stress necessary to move the dislocation at 0\,K. 

\subsection{Flexible boundary conditions}
\label{sec:cluster_flexible}

To remove the artifacts induced by the rigid boundary conditions, 
dislocation modeling with flexible boundary conditions has been developed.
The proposed method relies either on the use of the lattice Green's function \citep{Sinclair1978,Woodward2005}
or on the coupling with an empirical potential \citep{Liu2007,Chen2008}.

The lattice Green's function $G_{ij}(\vec{r})$
expresses, in the crystal harmonic approximation,  the displacement $\vec{u}$
induced on an atom in position $\vec{r}$ by a force $\vec{F}$ 
acting on an atom at origin:\footnote{We use the Einstein implicit summation 
convention on repeated indexes appearing in all expressions.}
\begin{equation}
	u_{i}(\vec{r}) = G_{ij}(\vec{r}) F_j .
	\label{eq:lattice_Green}
\end{equation}
This lattice Green's function can be obtained by inversion of the force-constant matrices 
of the perfect crystal \citep{Yasi2012,Tan2016} 
or can be tabulated from direct calculations in a perfect lattice
\citep{Sinclair1978,Rao1998}.
In the long range limit, $G_{ij}(\vec{r})$ converges to the elastic Green's function
given by anisotropic elasticity theory.

Flexible boundary conditions based on lattice Green's functions 
still make use of a cylinder geometry to model a single dislocation,
but three zones are now defined  (Fig. \ref{fig:cluster_sketch}b). 
Atoms in the inner zone (1) are relaxed with the \abinitio code until the forces acting on them 
are smaller than a fixed threshold, while atoms in zones (2) and (3) are kept fixed. 
At the end of this step, atomic forces have appeared in zone (2), because the dislocation elastic field
deviates from the Volterra solution used as an initial guess.
The lattice Green's function is then used to displace atoms in all three zones 
according to Eq. \ref{eq:lattice_Green} using all atomic forces in zone (2).
This leads to the cancellation of forces in zone (2) but makes new forces appear in zone (1). 
The procedure is thus iterated until all forces in zones (1) and (2) are null.
This self-consistent cycle is necessary because the lattice Green's function of the perfect crystal 
only approximates the linear response of the dislocated crystal.
Atoms in zone (3) serve as a buffer to prevent any perturbation by the external boundary
of forces building in zone (2).
As shown by \cite{Segall2003}, this buffer region may need to be quite large in metals
to obtain negligible perturbations in the inner regions.
This can be minimized by removing the surfaces delineating zone (3)
and using periodic boundary conditions in all directions.  
Interface defects are then present 
at the boundary between two periodic simulations cells.  But these defects 
lead to a smaller perturbation of the electronic density than the vacuum layer 
of the surfaces \citep{Woodward2005}.
One thus perfectly models an isolated dislocation in an infinite crystal
taking full account of the dislocation core field. 
It is possible to study dislocation cores with a reduced number of atoms 
in the simulation cell, a size usually compatible with \abinitio calculations.

A similar approach relies on the coupling of the \abinitio calculations with an empirical potential \citep{Liu2007,Chen2008}.
The simulation cell is still divided in three regions (Fig. \ref{fig:cluster_sketch}b). 
\Abinitio calculations are performed only for a smaller simulation cell corresponding to regions (1) and (2). 
Atoms in regions (2) and (3) are relaxed according to the forces calculated with the empirical potential,
whereas atoms in region (1) are relaxed according to \abinitio forces plus a correction to withdraw the perturbation 
caused by the external boundary of the \abinitio box.
The buffer region (2) has been added to the original approach \citep{Choly2005}
to minimize this correction by protecting atoms from the external boundary.
To operate, this method needs therefore an empirical potential which perfectly reproduces the lattice parameters 
given by \abinitio calculations, which can generally be done by rescaling the distances. 
Besides, the potential has also to match as best as possible the \abinitio linear response, 
\ie at least the elastic constants and, ideally, the whole phonon spectrum.

As it will be discussed in the last section, 
the main drawback of this \abinitio dislocation model using flexible boundaries
arises from the difficulty of extracting dislocation energy. 
The problem may be actually less sensitive with the second approach relying on a coupling with an empirical potential
where an energy formulation exists. In this case, one can obtain a reasonable estimation 
of the dislocation energy provided the potential gives an accurate description
of the boundary energy compared to the \abinitio calculations.
While these flexible boundaries truly allow the modeling of an isolated dislocation, 
thus predicting its core structure and its evolution under an applied stress 
without any \apriori artifact induced by the small size of the simulation cell 
inherent to \abinitio calculations, 
the approach is still under active development 
to also provide information on the dislocation energy.

\subsection{Periodic boundary conditions}

\begin{figure}[!b]
	\begin{center}
		\includegraphics[width=0.39\linewidth]{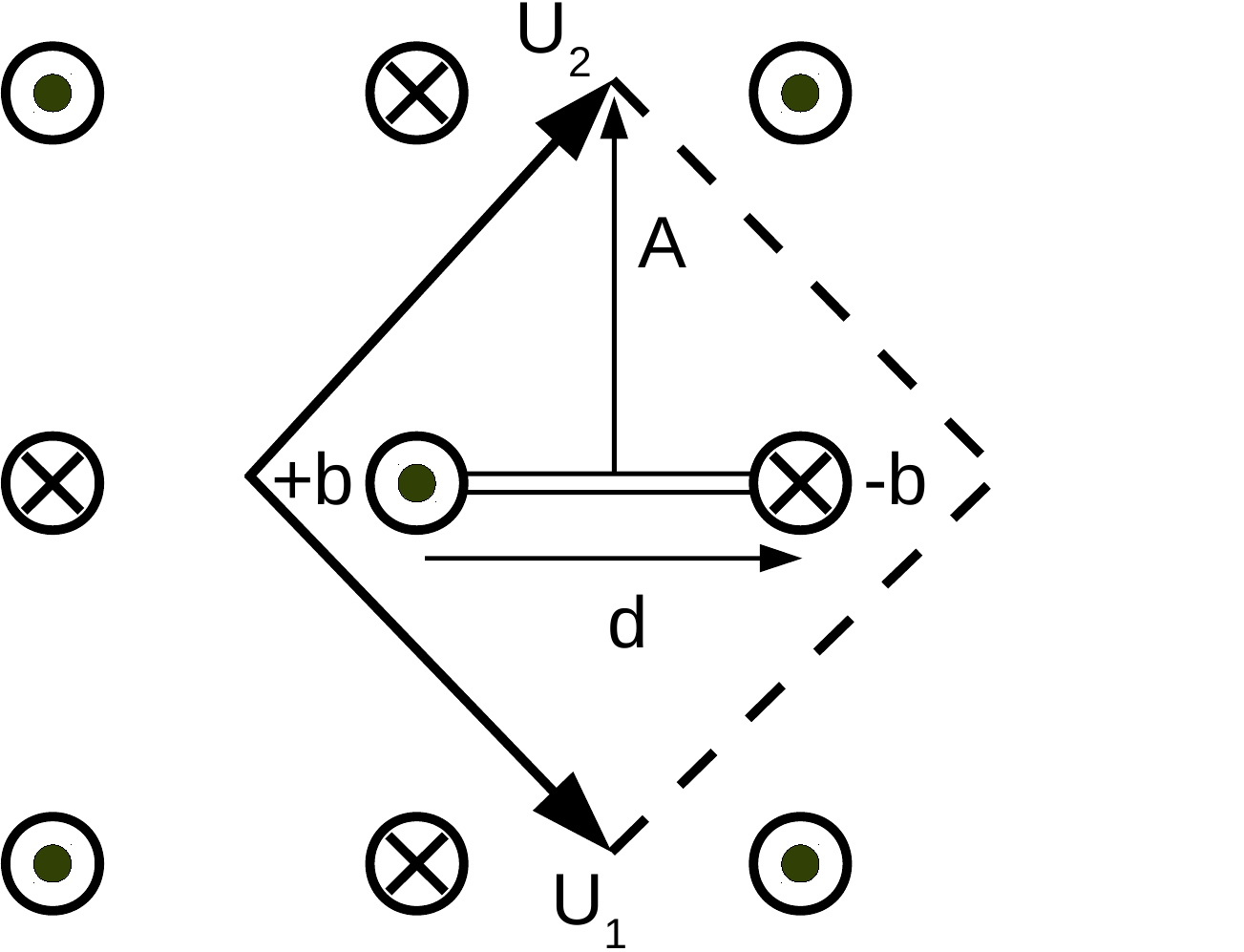}
		\hfill
		\includegraphics[width=0.6\linewidth]{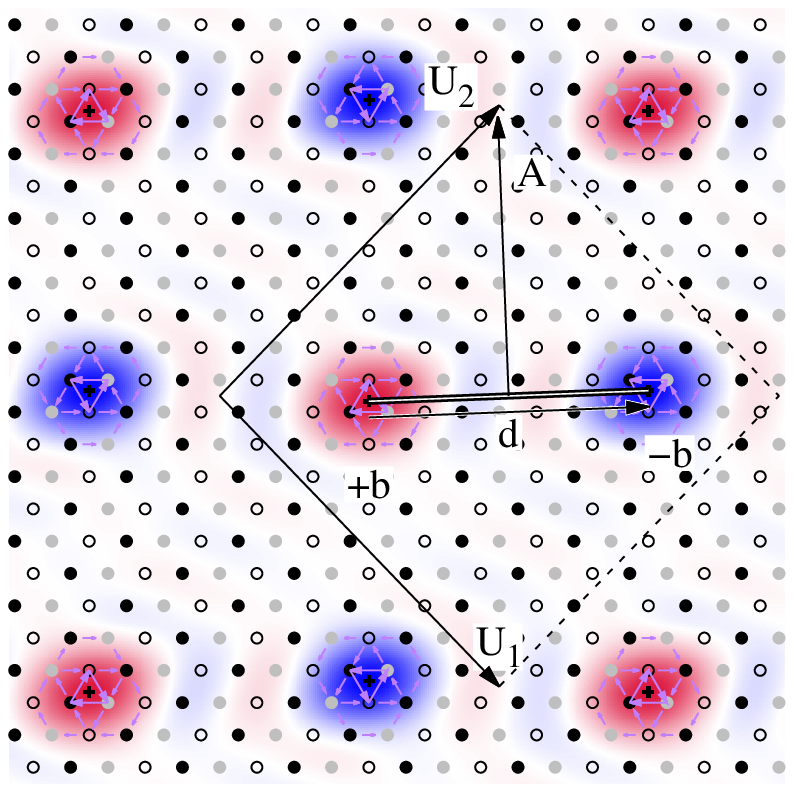}
	\end{center}
	\caption{Simulation of a dislocation dipole with periodic boundary conditions, 
	using a quadrupolar arrangement. 
	The dipole is defined by its Burgers vector $\vec{b}$,
	the dipole vector $\vec{d}$ joining the two dislocation centers, 
	and the cut vector $\vec{A}$,
	with the corresponding discontinuity surface indicated by a double black line.
	$\vec{u}_1$ and $\vec{u}_2$ are the periodicity vectors of the simulation cell
	perpendicular to the dislocation line.
	The example on the right corresponds to the simulation cell used for 
	the modeling of the $1/2\,\hkl<111>$ screw dislocation in bcc iron.
	The dislocation core structures are shown through their differential displacement maps 
	and their density (\cf Fig. \ref{fig:screw_core}a for a details).}
	\label{fig:PBC_sketch}
\end{figure}

To get rid of the external boundary and to use periodic boundary conditions
in all three directions without the introduction of a defective interface, one needs to introduce
in the simulation cell a dislocation dipole, \ie two dislocations with opposite Burgers vectors. 
One thus models a 2D periodic array of dislocations (Fig. \ref{fig:PBC_sketch}).

Several arrangements of dislocation arrays can be though off,  
but they are not all equivalent. 
Among all of them, the ones which are quadrupolar 
display strong advantages. 
A periodic array is quadrupolar, if the vector $\vec{d}$ linking the two 
dislocations of opposite signs is equal to $1/2\ (\vec{u}_1 + \vec{u}_2)$, 
where $\vec{u}_1$ and $\vec{u}_2$ are the periodicity vectors of the simulation cell
(Fig. \ref{fig:PBC_sketch}). This ensures that every dislocation is a symmetry center
of the array: 
fixing, as a convention, the origin at a dislocation center, if a dislocation $\vec{b}$ is located 
at the position $\vec{r}$, there will also be a dislocation $\vec{b}$ in $-\vec{r}$. 
The stress created by these two dislocations will cancel to first order at the origin
thanks to the symmetry property of the Volterra elastic field.\footnote{$\sigma_{\rm V}(-\vec{r})
= -\sigma_{\rm V}(\vec{r})$ with $\sigma_{\rm V}$ the Volterra stress field 
of a single dislocation.}
As a consequence, this quadrupolar periodic array minimizes the elastic interaction 
between the dislocations, and hence the Peach-Koehler force acting on each dislocation 
because of the image dislocations associated with periodic boundaries. 
It is the best-suited periodic array to extract dislocation core properties
from \abinitio calculations.

Linear elasticity is still used to build the initial configuration,
displacing all atoms according to the superposition of the displacement fields 
created by each dislocation composing the periodic array.
The summation on periodic images can be either performed in reciprocal space \citep{Daw2006}
or in direct space after regularization of the conditionally convergent sums \citep{Cai2003}.
The crystal orientation used in such elastic calculations should be chosen 
so as to fix the displacement discontinuity exactly in-between the two dislocations 
composing the dipole, thus preventing the propagation of this discontinuity to infinity.
The cut vector $\vec{A}$ defining this discontinuity (Fig. \ref{fig:PBC_sketch}) 
is therefore given by $\vec{A} = \vec{l} \times \vec{d}$, 
where $\vec{l}$ is the line vector of the dislocations 
and $\vec{d}$ the vector joining the centers of the $+\vec{b}$ dislocation 
to the $-\vec{b}$ one. 
If the scalar product $\vec{A}.\vec{b}$ is non-null, 
\ie if the dislocation dipole has an edge component 
and the displacement discontinuity does not coincide with the dislocation glide plane,
it is also necessary to insert atoms into or delete them from the original lattice 
at the discontinuity location, thus following the Volterra operation.

A homogeneous strain needs also to be applied to accommodate the plastic strain 
created by the dipole \citep{Daw2006,Cai2003} and ensure that the average stress
in the simulation cell is null. This can be easily demonstrated 
by considering the variation of the elastic energy when a homogeneous strain
$\varepsilon_{ij}$ is applied to a simulation cell containing a dislocation dipole
defined by its Burgers vector $\vec{b}$ and its cut vector $\vec{A}$
\begin{equation*}
	\Delta E(\varepsilon) = \frac{1}{2} S\,C_{ijkl}\,\varepsilon_{ij}\,\varepsilon_{kl}
	+ C_{ijkl}\,b_i\,A_j\,\varepsilon_{kl} ,
\end{equation*}
where energies are defined per dislocation unit length 
and have been thus normalized by the height of the simulation cell in the direction 
of the dislocation line.
$S$ is the area of the simulation cell perpendicular to this direction
and $C_{ijkl}$ are the elastic constants. 
The average stress existing in the simulation cell is then given by
\begin{equation}
	\sigma_{ij} = \frac{1}{S} \frac{\partial \Delta E}{\partial \varepsilon_{ij}}
	= C_{ijkl} \left( \varepsilon_{kl} - \varepsilon_{kl}^0 \right),
	\label{eq:PBC_stress}
\end{equation}
with the plastic strain defined by
\begin{equation}
	\varepsilon_{kl}^0 = - \frac{b_i A_j + b_j A_i}{2S}.
	\label{eq:PBC_plastic_strain}
\end{equation}
One thus sees that the stress given by Eq. \ref{eq:PBC_stress} is null
when the applied strain $\varepsilon_{ij}$ is equal to the plastic strain 
$\varepsilon_{ij}^0$.
When this applied strain is different, a Peach-Koehler force
acting on the dislocations may exist.
This allows studying properties of the dislocation core
under an applied stress, to determine its Peierls stress, for instance.
Finally, when a stress variation is observed in \abinitio calculations, 
Eq. \ref{eq:PBC_stress} allows to deduce the plastic strain from this stress,
and thus the dislocations' relative positions via the cut vector $\vec{A}$ (Eq. \ref{eq:PBC_plastic_strain}). 
For instance, the trajectories of the screw dislocations gliding between 
two neighbouring Peierls valleys have been determined thanks 
to this method in HCP Zr \citep{Chaari2014} and in BCC transition metals
\citep{Dezerald2016}.

With these periodic boundary conditions, all the excess energy contained 
in the simulation cell is due to the dislocations. 
As it will be shown in the last section, elasticity theory 
can be used to isolate the contribution of a single dislocation. 
These periodic boundary conditions offer thus a convenient way to extract 
dislocation energy from \abinitio calculations. 
But
the dislocation core structure, and hence the associated excess energy, 
can be perturbed by the presence of the periodic images. 
In practice, one will therefore need to check how sensitive 
are the obtained dislocation properties with the size of the simulation cell.

\section{Dislocation core structures}

Different representations can be used to image and analyse the relaxed dislocation core structure
obtained by atomic simulations. 
This allows, for instance, highlighting a spreading of or a dissociation of the dislocation.

\subsection{Differential displacement maps}

\begin{figure}[!b]
	\begin{center}
		\subfigure[bcc Fe: $\vec{b} = 1/2\,\hkl<111>$]{
			\includegraphics[height=0.22\textheight]{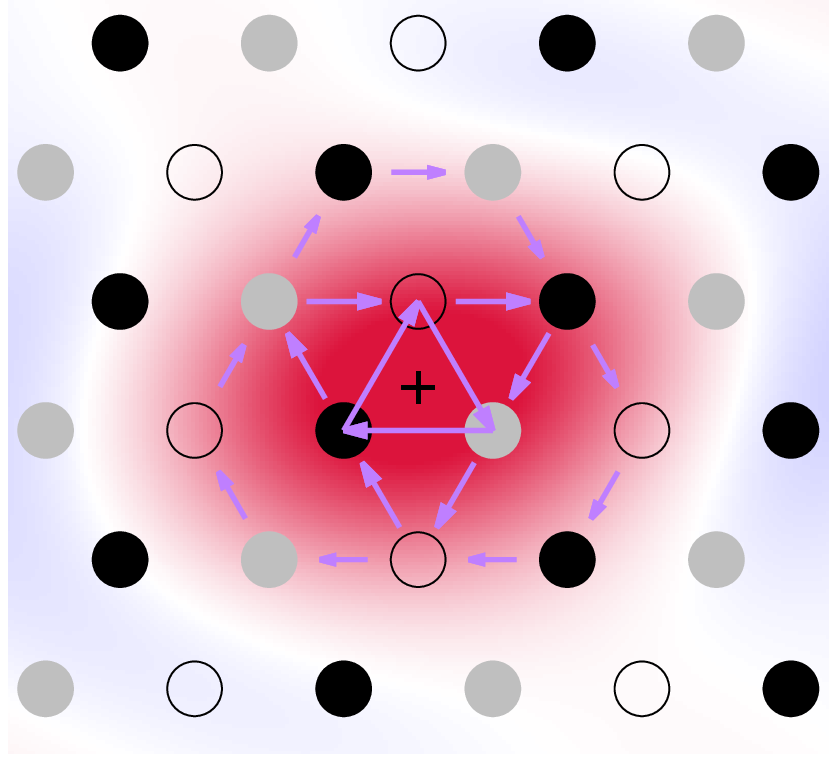}
			\includegraphics[height=0.20\textheight]{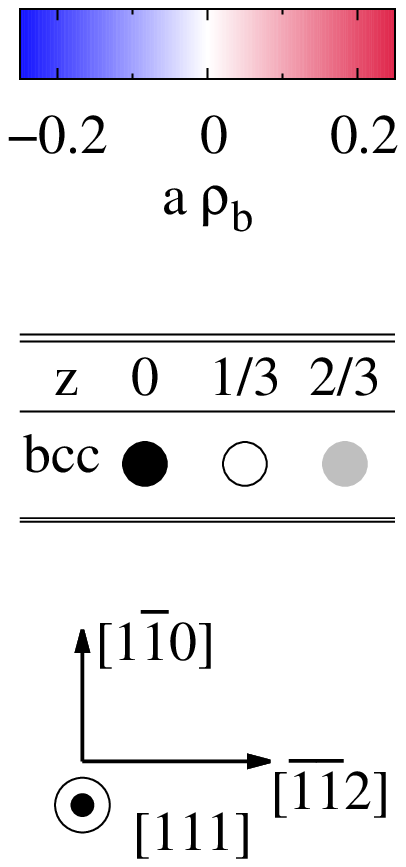} }
		\hfill
		\subfigure[hcp Zr: $\vec{b} = 1/3\,\hkl<1-210>$]{
			\includegraphics[height=0.25\textheight]{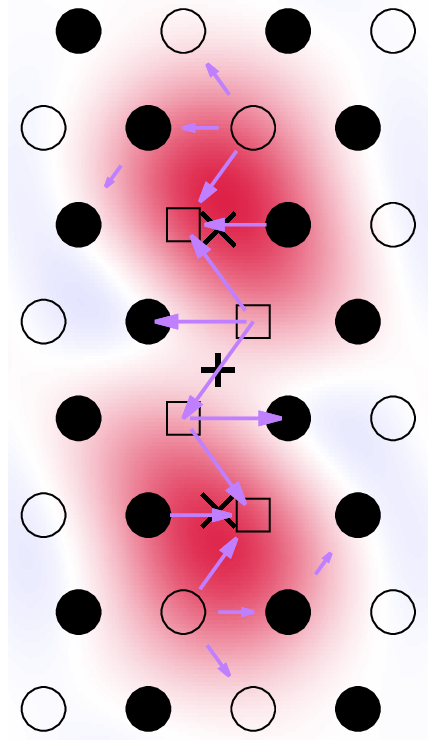}
			\includegraphics[height=0.20\textheight]{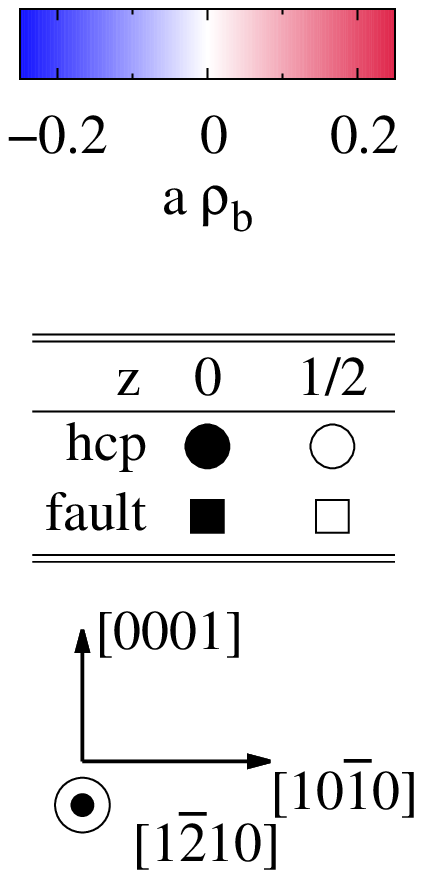} }
	\end{center}
	\caption{Core structure of a $\vec{b}$ screw dislocation (a) in bcc iron \citep{Dezerald2016}
	and (b) in hcp zirconium \citep{Clouet2015}.
	In these projections perpendicular to the dislocation line, 
	atoms are sketched by symbols with a colour depending on their (a) \hkl(111) and (b) \hkl(1-210)
	plane in the original perfect crystal. 
	In (b), different symbols are used for atoms depending on their neighbourhood in the dislocated crystal, 
	\ie close to the perfect hcp crystal (circles) or to the unrelaxed prismatic stacking fault (squares). 
	The arrows between atomic columns are proportional to the differential displacement created by the dislocation 
	in the direction of the Burgers vector. 
	The colour map show the dislocation density $\rho_{b}$ normalized by the lattice parameter (Nye tensor). 
	The center of the dislocation is indicated by a + cross.
	The $\times$ crosses in (b) correspond to the positions of the partial dislocations 
	deduced from the disregistry in Fig. \ref{fig:disregistry}.}
	\label{fig:screw_core}
\end{figure}

Differential displacement maps were introduced by \cite{Vitek1970}. 
Two examples are shown in Fig. \ref{fig:screw_core} for a screw dislocation 
in a body centered cubic (bcc) crystal and a hexagonal close packed (hcp) crystal.
In these maps, the crystal is projected in the plane perpendicular to the dislocation line, 
using for atomic columns the positions in the perfect crystal.
The differential displacement caused by the dislocation is calculated 
by considering the difference between the vector connecting two neighbour atoms
in the relaxed dislocated crystal and the same connecting vector in the perfect crystal. 
One then plots the projection of this differential displacement along the direction of the Burgers vector
with an arrow pointing from one atomic column to the other, 
centered in the middle of the two columns
and with an amplitude proportional to the differential displacement.
As the arrows are proportional to the displacement difference, 
they are a representation of the discrete derivative of the displacement field, 
\ie of the strain created by the dislocation.

The differential displacement map of the $1/2\,\hkl<111>$ screw dislocation in bcc Fe
shown in Fig. \ref{fig:screw_core}a highlights the compactness and the 3-fold symmetry of the core. 
Arrows have been normalized so that an arrow linking the centers of two atomic columns 
corresponds to a differential displacement of $b/3$.
One can thus draw Burgers circuits on this map and obtain the norm of the Burgers vector
of the enclosed dislocation by summing arrows.
The only non-null Burgers vector is obtained for circuits containing the dislocation center
indicated by a cross, 
in particular for the triangle connecting the three central \hkl[111] atomic rows, 
with a norm equal to $b$. 
The dislocation is thus well localized.

The picture is quite different for the $1/3\,\hkl<1-210>$ screw dislocation in hcp Zr
shown in Fig. \ref{fig:screw_core}b. 
The differential displacement map shows a non isotropic distribution 
with a spreading of the dislocation core in the \hkl(10-10) prismatic plane. 
The normalization here ensures that the maximal arrows correspond to a $b/2$ differential displacement.
The presence of a ribbon with arrows having almost the same length therefore corresponds to a $b/2$ prismatic stacking fault
which is known to be stable in this transition metal.
The differential displacement map thus clearly evidences the dissociation of the screw dislocation
in two $1/6\,\hkl<1-210>$ partial dislocations separated by a prismatic stacking fault.

\subsection{Dislocation density} 

Another visualization method proposed by \cite{Hartley2005} consists of extracting 
the Nye tensor from the relaxed atomic structure, thus giving a measure of the dislocation density.
The component $\alpha_{jk}$ of the Nye tensor corresponds to the density 
of dislocations with a line direction along $\vec{e}_k$
and a Burgers vector along $\vec{e}_j$.
If $A$ is a surface element of normal $\vec{n}$, 
the dislocation content of line defects along $\vec{n}$ 
intersecting $A$ is given by the surface integral
\begin{equation*}
	\vec{b} = \int_A{\alpha\cdot\vec{n}\,\ud{S}}.
\end{equation*}
We only give here the salient points of the method to extract the Nye tensor 
from atomic simulations
and the reader is referred to the original publication 
for the practical implementation. 

The first step is to define the elastic distortion, 
\ie the gradient of the elastic displacement, at each atomic position.  
Note that this differs from the gradient of the total displacement.
One cannot simply compare the atomic positions after and before the introduction of the dislocation
to obtain this elastic distortion, but one needs to find for each position 
the closest undistorted environment corresponding to a zero stress state.
This is performed by comparing, for each atom, the positions of its nearest neighbours 
in the dislocated relaxed crystal with the ones in a perfect crystal.  
Knowing the two sets of neighbour positions, each bond in the dislocated crystal,
defined by its vector $\vec{P}^{(\gamma)}$,
is identified with the corresponding $\vec{Q}^{(\beta)}$ bond in the perfect crystal,
which is the perfect bond leading to the smallest angle $\Phi^{(\gamma\beta)}$ 
between the vectors $\vec{P}^{(\gamma)}$ and $\vec{Q}^{(\beta)}$.
Only the bonds which are not too much distorted  and for which the angle $\Phi^{(\gamma\beta)}$
is smaller than a chosen threshold are kept. 
The elastic distortion $F^{\rm e}$ is then locally defined through the relation 
$P^{(\gamma)}_i = F^{\rm e}_{ij} \, Q^{(\beta)}_j$. 
This cannot be satisfied for each set of associated bond $(\gamma\,\beta)$ as the system of equations is overdetermined
and the matrix $F^{\rm e}$ is obtained by the pseudo-inverse method,
\ie a least-square fitting. 
The Nye tensor $\alpha$ is then defined through the rotational
of the inverse transpose of the distortion, 
$\alpha = - \nabla \times (^{+}F^{\rm e})^{-1}$,
using finite differences between neighbour atoms for derivation.
This defines the Nye tensor on a set of discrete points, generally atomic positions, 
which can be then interpolated with cubic-splines or Fourier series, 
or smeared with Gaussian-like spreading functions.

The dislocation density obtained for the $1/2\,\hkl<111>$ screw dislocation in bcc Fe 
(Fig. \ref{fig:screw_core}a) illustrates the compactness of the core: 
the distribution has only one peak. 
On the other hand,
the dislocation distribution for the $1/3\,\hkl<1-210>$ screw dislocation in hcp Zr
(Fig. \ref{fig:screw_core}b) shows two well-separated peaks which 
correspond to the two partial dislocations. 
To obtain the Nye tensor in this latter case, the neighbourhood of each atom 
in the dislocated crystal is compared not only to the two different neighbourhoods 
existing in the perfect hcp crystal, but also to the ones of the unrelaxed prismatic stacking fault, 
to identify the closer reference from which the elastic distortion is calculated.

\subsection{Disregistry}
\label{sec:disregistry}

\begin{figure}[!bh]
	\begin{center}
		\includegraphics[width=0.8\linewidth]{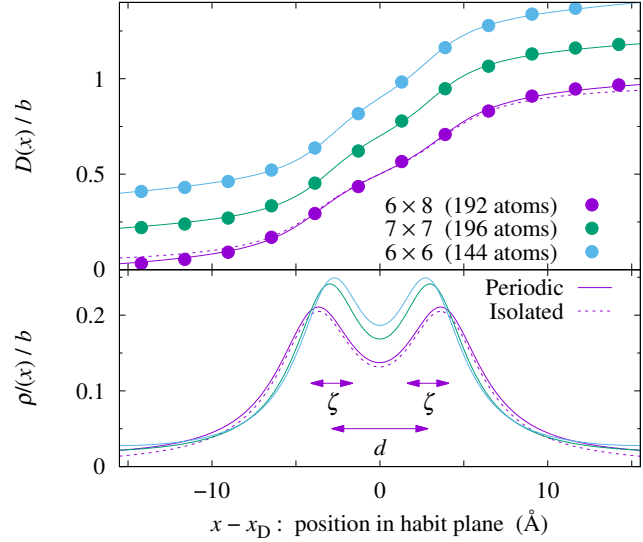}
	\end{center}
	\caption{Disregistry $D(x)$ created by a $1/3\,\hkl<1-210>$ screw dislocation 
	in its \hkl(10-10) prismatic glide plane in hcp Zr,
	and corresponding dislocation density $\rho(x)=\partial D(x)/\partial x$. 
	Symbols correspond to \abinitio calculations and lines to the fit of the Peierls-Nabarro model, 
	considering periodicity or not (straight and dashed lines respectively). 
	Results are shown for different $n \times m$ periodic arrangements 
	corresponding to the periodicity vectors $\vec{u}_1 = n/2\,\hkl[10-10]$
	and $\vec{u}_2 = m\,\hkl[0001]$ (see \cite{Clouet2012} for details).
	For clarity, disregistries $D(x)$ have been shifted by 0.2 between different data sets.
	The obtained dissociation distance $d$ and spreading $\zeta$ of the partial dislocations
	are indicated for the $6\times8$ periodic array whose core structure is shown in Fig. \ref{fig:screw_core}b.}
	\label{fig:disregistry}
\end{figure}

The extraction of the disregistry offers another way 
to characterize the dislocation core structure, 
particularly convenient when the core is planar. 
The disregistry is the difference of displacement induced by the dislocation
between the plane just above and the one just below the dislocation glide plane. 
It is thus obtained from the relaxed configuration through
\begin{equation*}
	\vec{D}(x) = \vec{u}^{+}(x) - \vec{u}^{-}(x),
\end{equation*}
where $\vec{u}^{+}(x)$ and $\vec{u}^{-}(x)$
are the displacements of the atoms belonging respectively to the upper and lower planes 
and located at the position $x$ in the direction perpendicular to the dislocation line.
This disregistry varies\footnote{If the cut plane used to introduce the dislocation 
in the simulation cell does not correspond to its glide plane,
it is necessary to define the atomic displacement in the $\vec{0}$
to $\vec{b}$ interval.  This can be done as the Burgers vector $\vec{b}$
of a perfect dislocation is a periodicity vector of the lattice 
and adding a displacement $n\vec{b}$ ($n \in \mathbb{Z}$) to an atom
does not change the configuration.}
from $\vec{0}$ for $x\to-\infty$ to $\vec{b}$ for $x\to\infty$,
thus corresponding to the dislocation glide plane
being locally sheared by one Burgers vector $\vec{b}$.
The dislocation center $x_{\rm D}$ is defined by $\vec{D}(x_{\rm D})=\vec{b}/2$.
The disregistry derivative, $\vec{\rho}(x) = \partial \vec{D}(x) / \partial x$,
corresponds to the dislocation density in the glide plane. 

Peierls and Nabarro built a model that leads to a simple analytical expression 
of the disregistry \citep{Lu2005}.
According to this model, the disregistry\footnote{For simplicity, 
we consider scalar quantities by projecting the displacement in the direction 
of the Burgers vector.} is given by 
\begin{equation*}
	D(x) = \frac{b}{\pi} \left[ \arctan{\left( \frac{x-x_{\rm D}}{\zeta} \right)} + \frac{\pi}{2} \right],
\end{equation*}
where $x_{\rm D}$ is the dislocation position and $\zeta$ its spreading in the glide plane.
Fitting of these two parameters to the data extracted from the atomistic simulations 
allows thus defining the dislocation position and characterizing the spreading of its core.

For dissociated dislocations, the disregistry is the sum of the contributions of the two partial dislocations,
\ie, assuming that each partial dislocation 
has the same Burgers vector $\vec{b}/2$ and the same spreading $\zeta$,
\begin{equation*}
	D(x) = \frac{b}{2\pi} \left[ \arctan{\left( \frac{x-x_{\rm D}-d/2}{\zeta} \right)} 
	+ \arctan{\left( \frac{x-x_{\rm D}+d/2}{\zeta} \right)} + \pi \right],
\end{equation*}
where $d$ is the dissociation distance. 
As shown in Fig. \ref{fig:disregistry} for the $1/3\,\hkl<1-210>$ screw dislocation in hcp Zr,
such an analytical expression generally perfectly describes the disregistry 
extracted from the atomic simulations. 
One can also notice that the positions in the glide plane of the partial dislocations 
deduced from the disregistry agree which what can be inferred 
from the differential displacement and the Nye tensor maps (Fig. \ref{fig:screw_core}b). 
Some consequences of the periodic boundary conditions used to model this dislocation 
are visible on these disregistry plots. 
The dislocation density slightly depends, through the dissociation distance $d$ 
and the partial spreading $\zeta$, on the simulation cell, not only its size but also its shape. 
One also sees that the density of the periodic dislocation arrays (solid line in Fig. \ref{fig:disregistry}),
obtained by summation of the contributions of the periodic images in the glide plane,
slightly differs from the one of the isolated dislocation (dashed line in Fig. \ref{fig:disregistry}),
especially in the distribution tail.
Flexible boundary conditions, as discussed in Section \ref{sec:cluster_flexible},
have been developed to solve such limitations 
of periodic boundary conditions.

\section{Dislocation energy}

\Abinitio calculations give access to the dislocation core energy
and its variations. 
This core energy is the part of the dislocation excess energy 
which arises from the strong perturbation of the atomic interactions 
in the immediate vicinity of the dislocation line and which cannot be described 
by linear elasticity. Contrary to the dislocation elastic energy, 
this is an intrinsic property which only depends on the dislocation 
and not on the surrounding environment. 
When several configurations exist for the same dislocation,
this core energy controls their relative  stability.
Its variations with the position of the dislocation in the crystal lattice
is at the origin of the lattice friction opposing dislocation glide.

\subsection{Core energy}

Among the different boundary conditions introduced in section \ref{sec:boundary} 
to model a dislocation at an atomic scale, 
only periodic boundary conditions allow for an unambiguous determination 
of the dislocation core energy with \abinitio calculations. 
This is a consequence of the energy formulation inherent to \abinitio calculations.
Because of the non-locality of the electronic energy, which contains a contribution 
which needs to be evaluated in reciprocal space, 
one cannot easily partition the excess energy of the simulation cell
between the dislocation and the external boundary contributions
when a defective boundary has been introduced 
like in cluster approaches using either fixed (\S \, \ref{sec:cluster_fixed})
or flexible boundaries (\S \, \ref{sec:cluster_flexible}). 
\Abinitio methods to project the energy on atoms have been proposed: 
they theoretically allow for such a partition but the application 
to the calculation of a dislocation core energy still remains to be done. 
Even if the absolute value of the core energy appears difficult to determine with cluster approaches,
methods to estimates its variation are nevertheless possible.
One can, for instance, calculate the difference of core energy between two configurations of the same dislocations
by simply considering the difference of \abinitio total energies. 
But such an approach assumes that the contribution of the external boundary will cancel in the difference,
an assumption which may be hard to validate.
Variation of the dislocation energy with its position in the crystal lattice 
can also be estimated by considering the work of the atomic forces during the motion
(\cite{Swinburne2017}).

On the other hand, with periodic boundary conditions, all the excess energy arises from the dislocations. 
This excess energy $\Delta E$ is defined as the energy difference per unit of height 
between the supercell with and without the dislocation dipole.\footnote{If atoms
have been removed or inserted during the creation of the dipole, 
the energy of the perfect supercell needs to be normalized by the correct
number of atoms.}
It is given by the sum of the core energy $E^{\rm core}$ of the two dislocations, 
of the elastic energy $E^{\rm elas}_{\rm dipole}$ of the dipole contained in the supercell
and of its elastic interaction with its periodic images:
\begin{equation}
	\Delta E = 2 \, E^{\rm core} 
	+ E^{\rm elas}_{\rm dipole}
	+ \frac{1}{2} \sum_{n,m}{ E^{\rm elas}_{\rm inter}(n\,\vec{u}_1 + m\,\vec{u}_2) }
	\label{eq:Edipole}.
\end{equation}
The factor $1/2$ appears in front of this last contribution 
as only one half of the interaction is attributed to each interacting dipole.
When partitioning the excess energy into a core and an elastic contribution,
it is necessary to introduce a cutoff distance to isolate the dislocation cores. 
Close to the dislocation lines, strains are much too high to be described by linear elasticity.
As a consequence, elastic fields diverge at the origin
and one needs to exclude the core region from the elastic description. 
The elastic contribution to the excess energy is thus obtained by integrating
the elastic energy density on the whole supercell except two cylinders of radius $r_{\rm c}$ 
which isolate this elastic divergence.
The core energy corresponds to the excess energy contained in these cylinders.
The total excess energy $\Delta E$ does not depend on the choice for this core radius, 
but the partition between a core and an elastic contribution depends on $r_{\rm c}$.

\begin{figure}[!b]
	\begin{center}
		\includegraphics[width=0.5\linewidth]{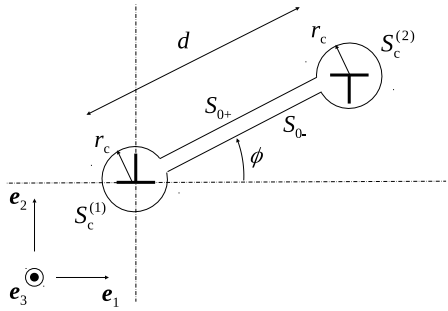}
	\end{center}
	\caption{Definition of the contour surface used to calculate 
	the elastic energy of a dislocation dipole}
	\label{fig:Eelas_integral}
\end{figure}

The elastic energy of the dipole and its interaction with its periodic images
can be computed by considering the Volterra elastic field created by the dislocations.
This calculation can be performed either in reciprocal space \citep{Daw2006} 
or in direct space using classical results of dislocation elastic theory \citep{Bacon1980}.  
In this last case, one uses the decomposition of Eq. \eqref{eq:Edipole}, 
with the contribution of the dipole contained in the supercell 
and its interaction with the periodic images calculated separately. 
The dipole elastic energy is obtained by the volume integral 
\begin{equation*}
  E^{\rm elas}_{\rm dipole} = \frac{1}{2}\iiint_V{
        \left( \sigma^{(1)}_{ij} + \sigma^{(2)}_{ij} \right)
        \left( \varepsilon^{(1)}_{ij} + \varepsilon^{(2)}_{ij} \right) \ud{V}},
\end{equation*}
where $\sigma^{(n)}$ and $\varepsilon^{(n)}$ are the stress and strain created by the dislocation $n$.
This is transformed into a surface integral thanks to Gauss' theorem
\begin{equation*}
        E^{\rm elas}_{\rm dipole} = \frac{1}{2} \iint_{S}{
                \left( \sigma_{ij}^{(1)} + \sigma_{ij}^{(2)} \right)
                \left( u_i^{(1)} + u_i^{(2)} \right) \ud{S_j} } ,
\end{equation*}
with $\vec{u}^{(n)}$ the displacement field associated with dislocation $n$.
The integration surface is composed of the two cylinders $S_{\rm c}^{(1)}$ 
and $S_{\rm c}^{(2)}$ of radii $r_{\rm c}$ 
removing the elastic divergence at the dislocation cores, 
and of the two surfaces $S_{0^-}$ and $S_{0^+}$ removing the 
displacement discontinuity along the dislocation cut (Fig.~\ref{fig:Eelas_integral}).
The integration on both core cylinders leads to the same contribution
\begin{equation}
	E^{\rm elas}_{\rm c}(\phi) = \frac{1}{2} \iint_{S_{\rm c}^{(1)}} \sigma_{ij}^{(1)} u_i^{(1)} \ud{S_j} 
	= \frac{1}{2} \iint_{S_{\rm c}^{(2)}} \sigma_{ij}^{(2)} u_i^{(2)} \ud{S_j} .
	\label{eq:Ec1}
\end{equation}
This contributions of the core tractions to the elastic energy \citep{Clouet2009a} 
should not be forgotten as it ensures that the elastic energy is a state variable 
compatible with the work of the Peach-Koehler forces.
Besides, in \abinitio calculations where the distance $d$ between the two dipole dislocations is small, 
this can lead to a non negligible contribution compared to the one associated with the integral 
along the cut surface, even for a screw orientation.
The elastic energy of the dislocation dipole is then
\begin{equation}
	E^{\rm elas}_{\rm dipole} = 2 E^{\rm elas}_{\rm c}(\phi) 
		+  b_i K_{ij}^0 b_j \ln{\left( \frac{d}{r_{\rm c}} \right)},
	\label{eq:Eelas_dipole}
\end{equation}
where the tensor ${\sf{\textbf K}}^0$ defined by \cite{Stroh1958,Stroh1962} only depends on the elastic constants.
The total elastic energy is finally obtained by adding the interaction 
of the dipole with its periodic images. 
But, one should realize that the summation on periodic images 
appearing in Eq. \eqref{eq:Edipole} is only conditionally convergent: 
it can be regularized with the method of \cite{Cai2003}.

\begin{figure}[!b]
	\begin{center}
		\includegraphics[width=0.7\linewidth]{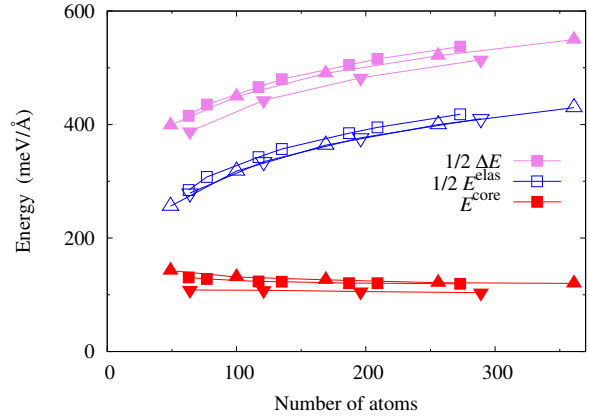}
	\end{center}
	\caption{Decomposition of the excess energy $\Delta E$ of a $1/2\,\hkl<111>$
	screw dislocation dipole in bcc Fe in an elastic contribution $E^{\rm elas}$
	and a core energy $E^{\rm core}$, using a core radius $r_{\rm c}=b/2$. 
	Different symbols correspond to different periodic arrangements 
	(see \cite{Clouet2009b} for details).}
	\label{fig:Ecore_screw_Fe}
\end{figure}

As shown in Fig. \ref{fig:Ecore_screw_Fe} for the $1/2\,\hkl<111>$
screw dislocation in bcc iron, once this elastic energy 
is subtracted from the dislocation excess energy given by \abinitio calculations,
one obtains a constant core energy which does not depend on the size of the supercell. 
Some slight variations of the core energy are nevertheless still observed 
with the type of periodic arrangement used for the atomic simulations. 
These variations arise because only the Volerra elastic field 
has been considered in the calculation of the elastic energy.
Dislocations also cause a core elastic field, which decays more rapidly 
than the Volterra elastic field.  
Because of the small size of the supercell used in \abinitio calculations, 
this core field may also lead to an elastic interaction between the different
dislocations composing the periodic array. 
This contribution to the elastic energy can be computed 
to improve the convergence of core energies \citep{Clouet2009b}. 
A quadrupolar periodic arrangement minimizes this contribution of the core field. 
This is why such an arrangement is preferred when periodic boundary conditions 
are used.
The neglect of anharmonic effects in the calculation of the elastic energy 
can also be a reason for the variation of the core energy with the supercell. 
Knowing higher order elastic constants, one can use non-linear elasticity theory in principle
to calculate more precisely this elastic contribution \citep{Teodosiu1982}.
But this leads to much cumbersome calculations. 
In practice, as anharmonicity is important only close to the dislocation core, 
the consideration of the dislocation core field offers a way 
to incorporate anhamonic effects while still relying on linear elasticity.

\subsection{Peierls energy barrier}
\label{sec:Peierls_barrier}

\begin{figure}[!b]
	a)\includegraphics[width=0.49\linewidth]{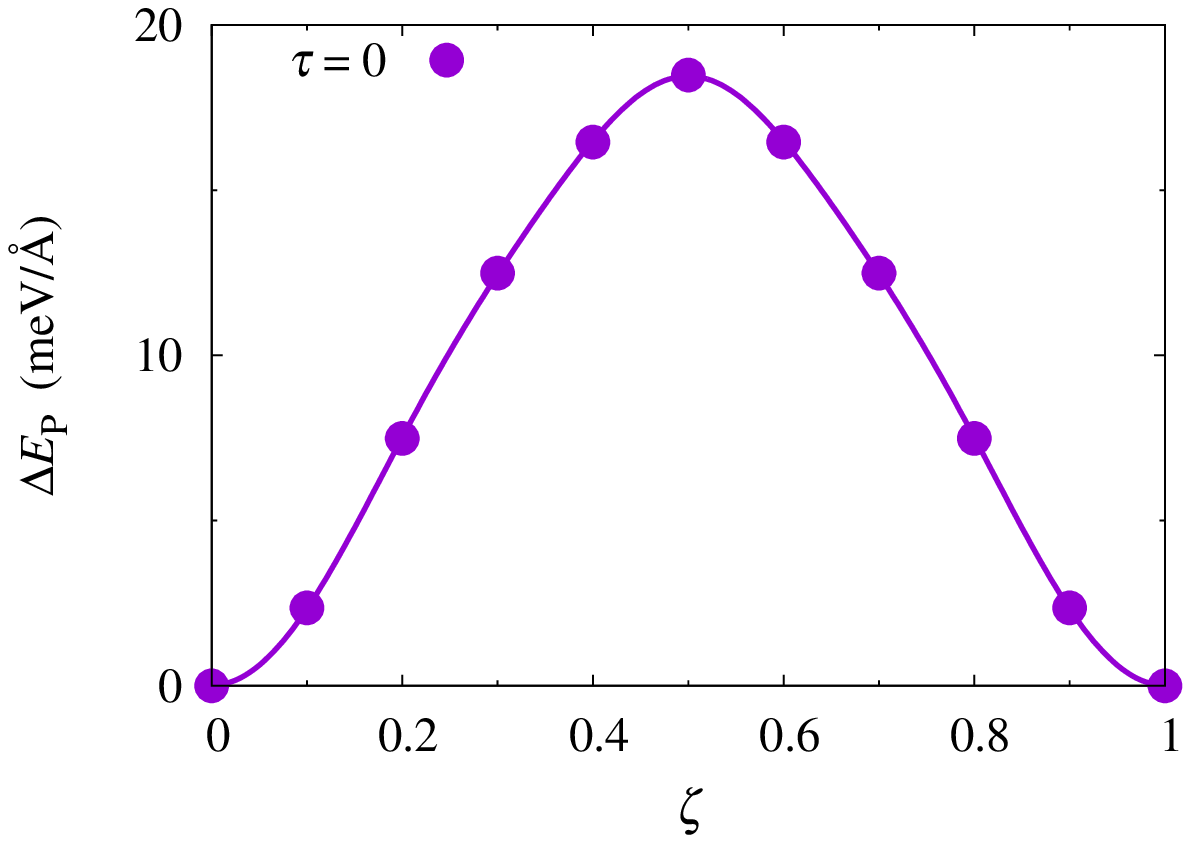}
	\hfill
	b)\includegraphics[width=0.49\linewidth]{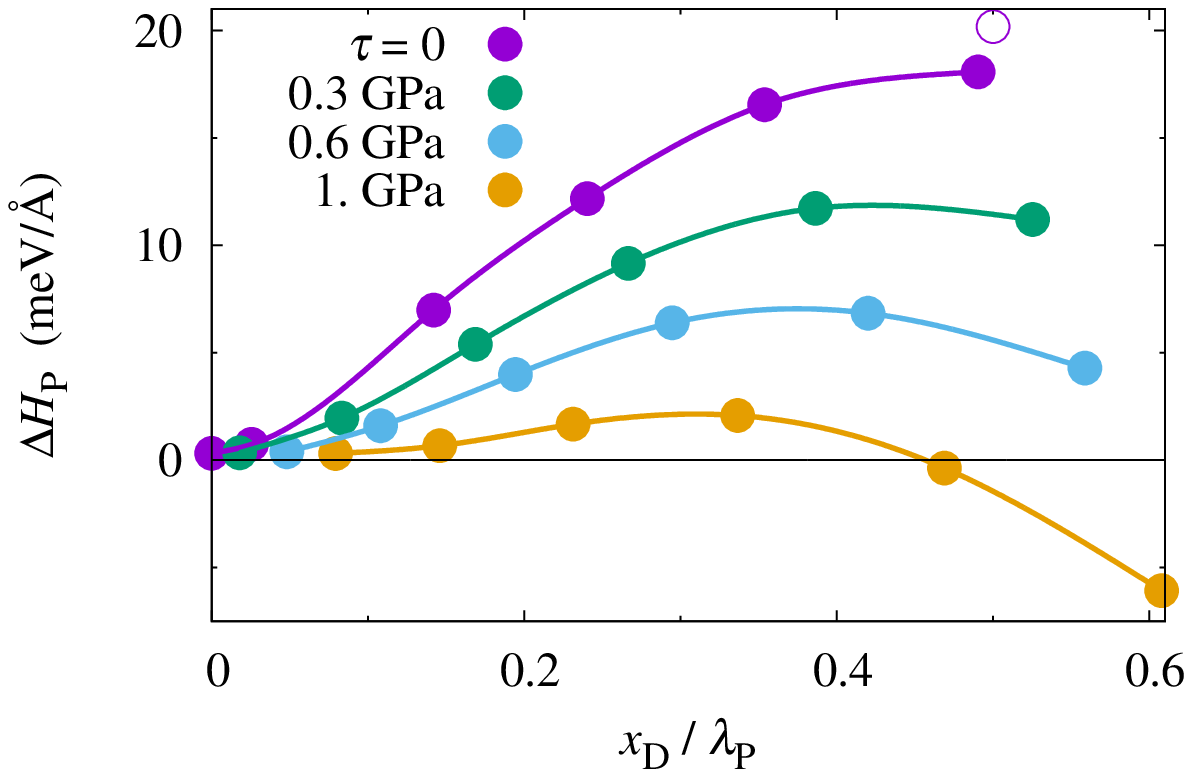}
	\caption{Peierls barrier of a $1/2\,\hkl<111>$ screw dislocation in bcc Mo.
	(a) The energy variation $\Delta E$ is shown as a function of the reaction coordinate $\zeta$,
	and (b) the enthalpy variation $\Delta H$ as a function of the dislocation position $x_{\rm D}$ normalized 
	by the distance $\lambda_{\rm P}$ between two Peierls valleys
	(see \cite{Dezerald2014,Dezerald2016} for details).
	For the Peierls barriers under stress (b),
	only one half of the barrier has been computed, 
	with one dislocation of the dipole being displaced while the second one remains fixed. 
	The open symbol is the enthalpy variation in the middle of the pathway ($x_{\rm D}/\lambda_{\rm P}=1/2$)
	before correcting for the variation of the elastic interaction energy for the $\tau=0$ calculation.}
	\label{fig:Peierls_Mo}
\end{figure}

The Peierls energy is the energy barrier opposing dislocation glide. 
It corresponds to a variation of the dislocation core energy 
as the elastic energy is not dependent upon the dislocation position in the crystal lattice.
It can be calculated by finding the minimum energy path linking two neighbouring stable positions 
of the dislocation using either constrained minimization or nudged elastic band (NEB) calculations \citep{Henkelman2000a}.

With periodic boundary conditions, the Peierls energy is directly obtained by considering 
a path where both dislocations composing the dipole are displaced by one Peierls valley
in the same direction.  
If the two dislocations are moved simultaneously along the path, 
their separation distance does not vary and the elastic energy is constant.  
This ensures that the energy variation given by the constrained minimization 
or the NEB calculations directly corresponds to the Peierls energy. 
However, this is possible only if crystal symmetry ensures that 
the path is symmetrical as the two dislocations are traversing 
their Peierls barriers in the opposite direction.  
This is the case, for instance, for the $1/2\,\hkl<111>$ screw dislocation 
in a bcc lattice gliding in a \hkl{110} plane (Fig. \ref{fig:Peierls_Mo}a).

If the path is not symmetrical, either because of the lack of crystal symmetries
or because of an applied stress, it is not possible anymore to move both dislocations
simultaneously in the same direction.  
One needs either to move them in opposite directions 
or to keep one dislocation fixed when the second one is moving. 
As a consequence, the separation distance, and thus the elastic interaction energy,
is varying along the path.  
One can calculate this variation of the elastic energy 
and subtract it from the excess energy in order to obtain 
the Peierls energy.  
To be able to perform this elastic calculation, one needs first
to determine the exact dislocation position $x_{\rm D}$
for each reaction coordinate $\zeta$ along the path.
This can be done using the dislocation disregistry (\cf \S \ref{sec:disregistry})
if the motion is planar or by fitting the atomic displacements
with the Volterra elastic solution.
As the stress is directly linked to the applied strain and the dislocation positions
(Eqs. \ref{eq:PBC_stress} and \ref{eq:PBC_plastic_strain}), 
one can also use the stress variation observed along the dislocation path 
to determine the dislocations position. 
The example of Fig. \ref{fig:Peierls_Mo}b shows that, with this correction for the variation 
of the elastic energy, the same Peierls energy is obtained under zero applied stress 
when one dislocation is fixed or when both dislocations are moved (Fig. \ref{fig:Peierls_Mo}a).

\subsection{Peierls stress}
\label{sec:Peierls_stress}

The Peierls stress is the applied resolved shear stress necessary to cancel the Peierls barrier
so that the dislocations can glide freely without the need of thermal activation,
\ie the stress necessary to move the dislocation at 0\,K.
For an applied stress $\tau$, the Peierls barrier is given by the enthalpy variation
\begin{equation*}
	\Delta H_{\rm P} (x_{\rm D}, \tau) = \Delta E_{\rm P} (x_{\rm D}) - \tau \, b \, x_{\rm D},
\end{equation*}
which corresponds to the Peierls energy barrier plus the work of the applied stress 
when the dislocation has glided a distance $x_{\rm D}$.
The Peierls stress is thus the maximum applied stress $\tau$ for which the function 
$\Delta H_{\rm P} (x_{\rm D}, \tau)$ 
goes through a maximum in the range 
$0 \leq x_{\rm D} \leq \lambda_{\rm P}$. 
If one assumes that the energy barrier $\Delta E_{\rm P} (x_{\rm D})$ does not depend on the applied stress $\tau$, 
it is given by
\begin{equation}
	\tau_{\rm P} = \frac{1}{b} \Max{\left( \frac{\partial \Delta E_{\rm P}}{\partial x_{\rm D}} \right)}.
	\label{eq:Peierls_stress}
\end{equation}
The Peierls stress can thus be theoretically obtained from the calculation of the Peierls energy barrier 
under zero applied stress.  
But, the evaluation of the derivative in Eq. \ref{eq:Peierls_stress} 
requires to know the variation of the energy as a function of the dislocation position
and not only of the reaction coordinate.
In practice, the obtained value for $\tau_{\rm P}$ will sensitively vary with the method chosen 
to estimate the dislocation position along the path. 
 
One can also directly calculate with \abinitio calculations the Peierls barrier under an applied stress
so as to estimate at which stress the barrier cancels (Fig. \ref{fig:Peierls_Mo}).
In such calculations, one does not really apply a stress
but a strain corresponding to the target stress (Eq. \ref{eq:PBC_stress}). 
With periodic boundary conditions, as the distance between 
the two dislocation is varying, the applied stress is also varying along the path.
Eqs. \eqref{eq:PBC_stress} and \eqref{eq:PBC_plastic_strain} show that the stress variation
is directly proportional to the dislocation displacement
and to the inverse of the surface $S$ of the simulation cell 
perpendicular to the dislocation line. 
If only one dislocation is moving along the path, 
this stress variation therefore does not exceed  
\begin{equation*}
	\delta \tau = \mu \frac{b \lambda_{\rm P}}{S},
\end{equation*}
where $\mu$ is the shear modulus in the dislocation glide plane.

If one is only interested in the calculation of the Peierls stress 
and not in the variation of the Peierls barrier with the applied stress, 
one can simply perform static relaxation of a dislocation under an applied stress 
to see at which applied stress the dislocation glides by at least one Peierls valley. 
With periodic boundary conditions one still needs to take into account 
the variation of the elastic interaction and of the applied stress 
when the dislocation is moving to interpret the results. 
On the other hand, no such artifact exists with a cluster approach 
using flexible boundary conditions which truly models a single isolated dislocation
under an applied stress. 
Straining homogeneously the simulation cell 
to obtain the targeted applied stress, the Peierls stress 
is defined as the stress for which the dislocation cannot be stabilized anymore
and escapes from the cluster.
If one is only interested in the evolution of the dislocation core structure 
under an applied stress and on the determination of the Peierls stress, 
this cluster approach therefore appears as the method of choice.
Nevertheless, whatever the boundary conditions, determination of the Peierls stress 
by such an instability condition of the dislocation core under an applied stress 
necessitates a strict threshold criterion on the atomic forces
to obtain a meaningful value.

\section{Conclusions}

Dislocation core properties can now be routinely studied with \abinitio calculations
thanks to the different methodological developments summarized in this chapter.  
This usually necessitates a coupling between atomistic model and elasticity theory, 
for which different already available tools can be used: 
see, for instance, D. R. Trinkle website\footnote{\url{http://dtrinkle.matse.illinois.edu}}
for an implementation of the lattice Greens functions 
or the Babel package\footnote{\url{http://emmanuel.clouet.free.fr/Programs/Babel}} for handling dislocations 
in atomistic simulation cells and elastic energy calculations.
Useful information on the dislocation core structure are thus obtained. 
Such calculations can, for instance, characterize possible dissociation or spreading of the core, 
or evidence the existence of several stable configurations for the same dislocation.
One gets access to the different energy barriers opposing the dislocation motion 
and to their variation with the applied stress. 
It is also possible to study how these core properties are altered by the interaction 
with solute atoms. 

Because of the limited size that can be handled by \abinitio calculations, 
such studies are usually limited to the study of straight dislocation, 
and only few \abinitio calculations have considered until now the presence of kinks on the dislocation lines. 
Upscaling modeling approaches, relying, for instance, to the line tension approximation 
to describe kink nucleation,
are therefore needed to go from these fundamental core properties 
determined at 0\,K with \abinitio calculations to dislocation mobility laws 
at finite temperature.  
Larger atomistic simulations are also possible using empirical potentials 
to describe atomic interactions.  These simulations allow studying 
more complex situations 
and simulating different dislocation mechanisms, 
like glide, cross-slip and interaction with other elements of the microstructure,
without assuming \apriori the elementary mechanism.
In such a context, \abinitio calculations are useful to validate 
and also help the development of empirical potentials 
which correctly reproduce dislocation fundamental properties.


\begin{acknowledgement}
	Drs. Nermine Chaari, Lucile Dezerald, and Lisa Ventelon are acknowledged for their contributions 
	to the works presented here. 
	Dr. Antoine Kraych is thanked for fruitful discussions.
	Parts of this work have been performed using HPC resources from GENCI-CINES and -TGCC (Grant 2017-096847).
\end{acknowledgement}

\bibliographystyle{spbasic}
\bibliography{clouet2018} 

\end{document}